\documentclass[conference]{IEEEtran}
\usepackage{hyperref}
\usepackage{breakurl}
\usepackage{bookmark}
\usepackage{subfig}
\usepackage{graphicx}
\usepackage{algpseudocode}
\usepackage{mathtools}
\usepackage{amsmath}
\usepackage{amsfonts}
\usepackage{amssymb}
\usepackage{mathptmx}
\usepackage{bbm}
\usepackage{multirow}
\usepackage{pifont}
\usepackage{sidecap}
\usepackage{examplep}
\usepackage[bb=boondox]{mathalfa}
\usepackage[capitalise]{cleveref}
\graphicspath{{./img/}}
\newcommand{\FLOP}{\mbox{flop}}
\newcommand{\BYTES}{\mbox{bytes}}
\newcommand{\TFLOPS}{\mbox{Tflop/s}}
\makeatletter
\presetkeys{todonotes}{inline}{}
\let\OldStatex\Statex
\renewcommand{\Statex}[1][3]{%
      \setlength\@tempdima{\algorithmicindent}%
        \OldStatex\hskip\dimexpr#1\@tempdima\relax}
        \makeatother

\newcommand{\ii}{\text{i}}
\newcommand{\ee}{\text{e}}

\begin{document}
\title{Performance Engineering of the Kernel Polynomial Method on Large-Scale
CPU-GPU Systems}

\author{\IEEEauthorblockN{Moritz Kreutzer, Georg Hager, Gerhard Wellein}
\IEEEauthorblockA{Erlangen Regional Computing Center\\
Friedrich-Alexander University of Erlangen-Nuremberg\\
Erlangen, Germany\\
\{moritz.kreutzer, georg.hager, gerhard.wellein\}@fau.de}
\and
\IEEEauthorblockN{Andreas Pieper, Andreas Alvermann, Holger Fehske}
\IEEEauthorblockA{Institute of Physics\\
Ernst Moritz Arndt University of Greifswald\\
Greifswald, Germany\\
\{pieper, alvermann, fehske\}@physik.uni-greifswald.de}
}

\maketitle

\begin{abstract}
The Kernel Polynomial Method (KPM) is a well-established scheme in quantum
physics and quantum chemistry to determine the eigenvalue density and spectral
properties of large sparse matrices. In this work we demonstrate the high
optimization potential and feasibility of peta-scale heterogeneous CPU-GPU
implementations of the KPM. At the node level we show that it is possible to
decouple the sparse matrix problem posed by KPM from main memory bandwidth both
on CPU and GPU. To alleviate the effects of scattered data access we combine
loosely coupled outer iterations with tightly coupled block sparse matrix
multiple vector operations, which enables pure data streaming. All optimizations
are guided by a performance analysis and modelling process that indicates how
the computational bottlenecks change with each optimization step. Finally we use
the optimized node-level KPM with a hybrid-parallel framework to perform large
scale heterogeneous electronic structure calculations for novel topological
materials on a petascale-class Cray XC30 system.
\end{abstract}


\IEEEpeerreviewmaketitle
It is widely accepted that future supercomputer architectures will change
considerably compared to the machines used at present for large scale
simulations. Extreme parallelism, use of heterogeneous compute devices and a
steady decrease in the architectural balance in terms of main memory bandwidth
vs.\ peak performance are important factors to consider when developing and
implementing sustainable code structures. Accelerator-based systems
already account for a performance share of 34\% of the total
TOP500~\cite{TOP500} today, and they may provide first blueprints of future
architectural developments. The heterogeneous hardware structure typically calls
for a completely new software development, in particular if the simultaneous use of all
compute devices is addressed to maximize performance and
energy efficiency.  

A prominent example demonstrating the need for new software implementations and
structures is the MAGMA project~\cite{MAGMA}. In dense linear algebra the
code balance (\BYTES/\FLOP) of basic operations 
can often be reduced by blocking techniques to better match the machine balance. 
Thus, this community is expected to achieve high absolute performance also on future
supercomputers. In contrast, sparse linear algebra is known for low sustained
performance on state of the art homogeneous systems. The sparse matrix vector
multiplication (SpMV) is often the performance-critical step. Most of the broad
research on optimal SpMV data structures has been devoted to drive the balance
of a general SpMV (not using any special matrix properties) 
down to its minimum value of $6\,\BYTES/\FLOP$ (double precision)
or $2.5\,\BYTES/\FLOP$ (double complex) on all architectures, which is still at least an order of
magnitude away from current machine balance numbers. Just recently the long
known idea of applying the sparse matrix to multiple vectors at the same time
(SpMMV) (see, e.g., \cite{Gropp99}), to reduce computational balance has gained
new interest \cite{Liu12,Aktulga14}. 

A representative of the numerical sparse linear algebra schemes used in 
applications that can benefit from SpMMV is the Kernel Polynomial
Method (KPM). KPM was originally devised for the computation of eigenvalue
densities and spectral functions~\cite{SRVK96}, and soon found applications
throughout physics and chemistry (see~\cite{Weisse06} for a review). KPM can be
broadly classified as a polynomial-based expansion scheme, with the
corresponding simple iterative structure of the basic algorithm that addresses
the large sparse matrix from the application exclusively through SpMVs.  Recent
applications of KPM include, e.g., eigenvalue counting for predetermination of
sub-space sizes in projection-based eigensolvers~\cite{dNPS13} or for large
scale data analysis~\cite{BIBC13}. 

In this paper we present for the first time a structured performance
engineering process for the KPM that substantially brings down the
computational balance of the method, leading to high sustained
performance on CPUs and GPUs. The algorithm itself is untouched; all
optimizations are strictly changes to the implementations. We apply
a data-parallel approach for combined CPU-GPU parallelization and
present the first large-scale heterogeneous CPU-GPU computations for
KPM.  The main contributions of our work which are of broad interest
beyond the original KPM community are as follows:

We achieve a systematic reduction of code balance for a widely used
sparse linear algebra scheme by implementing a tailored, algorithm-specific
(``augmented'') SpMV routine instead of relying on a series of sparse linear 
algebra routines taken from an optimized general library like BLAS. We
reformulate the algorithm to use SpMMV in order to combine loosely coupled outer
iterations.
Our systematic performance analysis for the SpMMV operation on both
CPU and GPU indicates that SpMMV decouples from main memory
bandwidth for sufficiently large vector blocks, and that data cache access
then becomes a major bottleneck on both architectures.
Finally we demonstrate the feasibility of large-scale CPU-GPU KPM computations
for a technologically highly relevant application scenario, namely topological materials.
In our experiments, the augmented SpMMV KPM version achieves more than 
$100\,\TFLOPS$ on 1024 nodes of a CRAY XC30 system. This is equivalent to almost 10\% of
the aggregated CPU-GPU peak performance.

An open-source program library containing all presented software developments as well as the KPM
application code are available for download~\cite{GHOST}.
 


\subsection{Related Work}
SpMV has been -- and still is -- a highly active subject of research due to
the relevance  of this operation in applications of computational science
and engineering. It has turned out that the sparse matrix storage format is a
critical factor for SpMV performance. Fundamental research on sparse matrix
formats for the architectures considered in this work has been conducted by
Barrett et al.~\cite{Barrett94} for cache-based CPUs and Bell et al.~\cite{Bell09}
for GPUs. The assumption that efficient sparse matrix storage formats are
dependent on and exclusive to a specific architecture has been refuted by Kreutzer et
al.~\cite{Kreutzer14} by showing that a unified format (SELL-C-$\sigma$)
can yield high performance on both architectures under consideration in this
work. Vuduc~\cite{Vuduc03} provides a comprehensive overview of optimization
techniques for SpMV. 

Early research on performance bounds for SpMV and SpMMV has been done by Gropp
et al.~\cite{Gropp99} who established a performance limit taking into
account both memory- and instruction-boundedness. A similar approach has been
pursued by Liu et al.~\cite{Liu12}, who established a finer yet similar performance model for
SpMMV. Further refinements to this model have been accomplished by Aktulga et
al.~\cite{Aktulga14}, who not only considered memory- and
instruction-boundedness but also bandwidth bounds of two different cache levels.

On the GPU side, literature about SpMMV is scarce. The authors of ~\cite{Bell09}
mention the potential performance benefits of SpMMV over SpMV in the outlook of
their work. 
Anzt et al.~\cite{Anzt14} have recently presented a GPU implementation of SpMMV together with
performance and energy results. The fact that SpMMV is implemented in
the library cuSPARSE~\cite{cuSPARSE}, which is shipped together with the CUDA
toolkit, proves the relevance of this operation.

Optimal usage patterns for heterogeneous supercomputers have become an
increasingly important topic with the emergence of those architectures. An
important attempt towards high performance heterogeneous execution is
MAGMA~\cite{MAGMA}. However, the hybrid functions delivered by this toolkit are
restricted to dense linear algebra. Furthermore, MAGMA employs task-based work distribution, 
in contrast to the symmetric data-parallel approach used in this work.
Matam et al.~\cite{Matam12} have implemented
a hybrid CPU/GPU solver for sparse matrix-multiplication. However,
they do not scale their solution beyond a single node.

Zhang et al.~\cite{Zhang13} have presented a KPM implementation for a
single NVIDIA GPU, but they do not follow the conventional data-parallel 
approach. Memory footprint and corresponding main memory
access volume of their implementation scale linearly with the number
of active CUDA blocks, which limits applicability and performance severely.


\subsection{Application Scenario: Topological Materials}
To support our performance analysis with benchmark data from a real application we will apply our improved KPM implementation to a problem of current interest,
the determination of electronic structure properties of a three-dimensional (3D) topological insulator.

Topological insulators form a novel material class similar to graphene with promising applications in fundamental research and technology~\cite{Hasan10}.
The hallmark of these materials is the existence of topologically conserved quantum numbers, which are related to the familiar winding number from two-dimensional geometry, or to the Chern number of the integer quantum Hall effect.
The existence of such strongly conserved quantities makes topological materials first-class candidates for quantum computing and quantum information applications.

The theoretical modelling of a typical topological insulator is specified by the Hamilton operator
\begin{equation}\label{Ham}
  \begin{split}
         H = -t \sum_{\substack{j=1,2,3 \\[0.2ex] n}} \left(  \Psi_{n+\hat \ee_j}^\dagger
                  \frac{\Gamma^1 - \ii \Gamma^{j+1} }{2}
                \Psi_{n} + \text{H.c.}\right)  
   \\
              + \sum_{n} \Psi_{n}^\dagger \left( 
                      V_n     \Gamma^0 
                    + 2       \Gamma^1
                   \right) {\Psi_{n}} \;,
   \end{split} 
\end{equation}
which describes the quantum-mechanical behavior of an electric charge in the material, subject to an external electric potential $V_n$ that is used to create a superlattice structure of quantum dots.
The vector space underlying this operator can be understood as the product of a local orbital and spin degree of freedom, which is associated with the $4 \times 4$ Dirac matrices $\Gamma^a$, and the positional degree of freedom $n$ on the 3D crystalline structure composing the topological insulator.
We cite the Hamilton operator for the sake of completeness although its precise form is not relevant for the following investigation.
For further details see, e.g., Refs.~\cite{Schubert11,Pieper14}.

From the expression~\eqref{Ham} one obtains the sparse matrix representation of the Hamilton operator by choosing appropriate boundary conditions and spelling out the entries of the matrices $\Gamma^a$.
Here, we treat finite $N_x \times N_y \times N_z$ samples,
such that the matrix  $H$ in the KPM algorithm
has dimension $N = 4 N_x \times N_y \times N_z$.
The matrix is complex and Hermitian,
the number of non-zero entries is $N_\mathrm{nz} \approx 13 N$.

Characteristic for these applications is the presence of several sub-diagonals
in the matrix.
Periodic boundary conditions in the $x$ and $y$ directions lead to outlying diagonals in the matrix corners.
In the present example, the matrix is a stencil but not a band matrix.
Because of the quantum dot superlattice structure, translational symmetry is not available to reduce the problem size.
This makes the current problem relevant for large-scale computations.

One basic quantity of interest for physics applications is the eigenvalue density, or density of states (DOS),
\begin{equation}\label{DOS}
 \rho(E) = \sum_{n=1}^N \delta(E-E_n) = \mathrm{tr} [\delta(E\mathbb{1}  - H) ] \;,
\end{equation}
where the sum of the trace $\mathrm{tr}[\dots]$ runs over all eigenvalues $E_n$ of $H$.
The DOS quantifies the number of eigenvalues per interval,
and can also be used, e.g., to predict the required size of sub-spaces for eigenvalue projection techniques~\cite{dNPS13,pamm:2014}.

\begin{figure}[t]
\centering
\includegraphics[height=\columnwidth,angle=270,origin=rb]{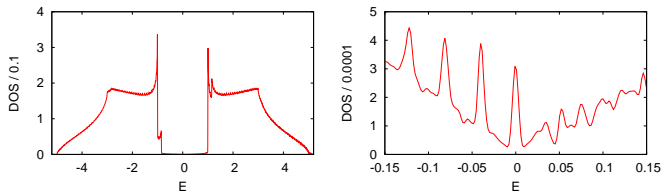}
\caption{DOS for a $1600 \times 1600 \times 40$ topological insulator ($N \approx 4 \times 10^8$) computed
with the KPM-DOS algorithm.}
\label{fig:topi_dos}
\end{figure}

\begin{figure}[t]
\centering
\includegraphics[height=3.6cm,angle=270,origin=cb]{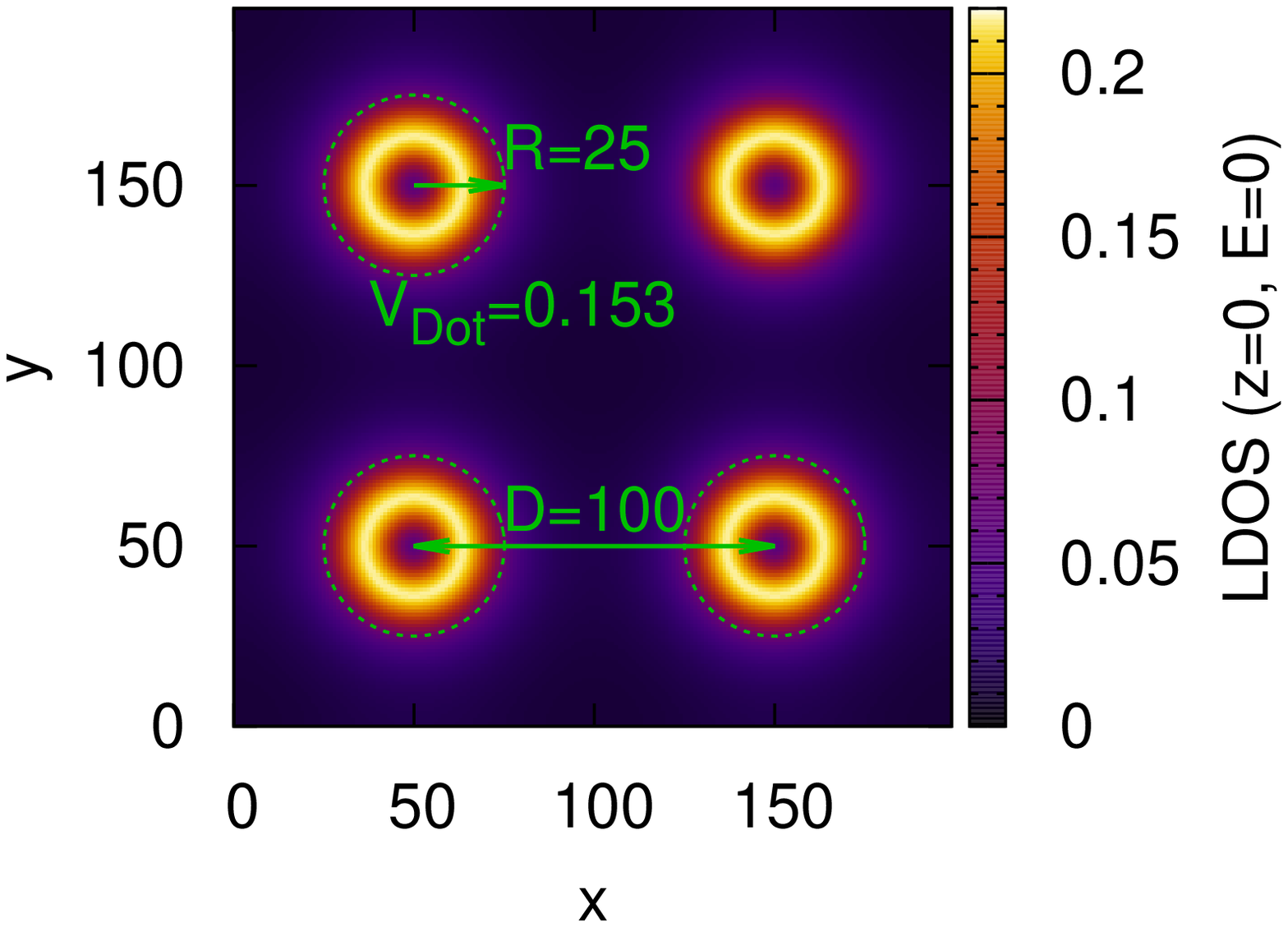}
\hspace{1mm}
\includegraphics[height=4.6cm,angle=270,origin=cb]{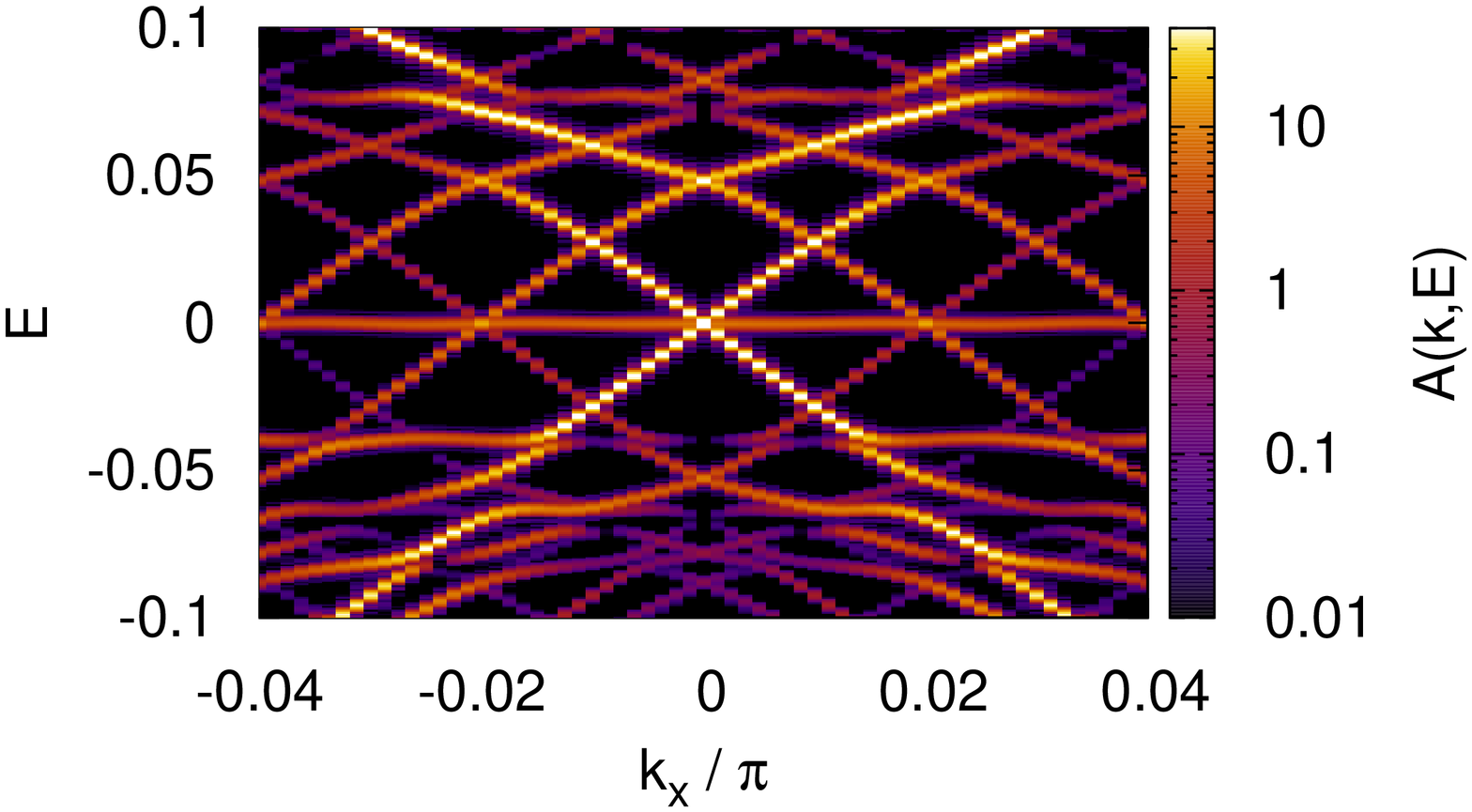}
\caption{Left panel: Local DOS for a quantum dot superlattice imposed on top of a topological insulator. Right panel: Corresponding momentum-resolved spectral function $A(k,E)$.
See, e.g., Refs.~\cite{Schubert11,Pieper14} for details on the physics.}
\label{fig:topi_demo}
\end{figure}

A direct method for computation of $\rho(E)$ that uses the first expression in \eqref{DOS} would have to determine all eigenvalues of $H$, which is not feasible for large matrices.
Instead, we rely on the KPM-DOS algorithm introduced in the next section.
In Figs.~\ref{fig:topi_dos},~\ref{fig:topi_demo} a few data for the DOS obtained with KPM-DOS are shown for the present application.


\section{Problem Description}
\label{sec:kpm}
The KPM is a polynomial expansion technique for the computation of spectral quantities of large sparse matrices (see our review~\cite{Weisse06} for a detailed exposition).
In the context of the definition \eqref{DOS} the KPM does not work directly with the first expression, but with a systematic expansion of the $\delta$-function in the second expression.
KPM is based on the orthogonality properties and two-term recurrence for Chebyshev polynomials $T_m(x)$ of the first kind.
Specifically, in KPM one successively computes the vectors $|\nu_m\rangle = T_m(\tilde H) |\nu_0 \rangle$ from a starting vector $|\nu_0\rangle$, for  $1 \le m \le M/2$ with prescribed $M$, through the recurrence
\begin{equation} \label{eq:kpm_iter}
| \nu_1 \rangle = \tilde H | \nu_0 \rangle \;, \quad
| \nu_{m+1} \rangle = 2 \tilde H | \nu_m \rangle - | \nu_{m-1} \rangle \;.
\end{equation}
The recurrence involves the matrix $\tilde H$ only in (sparse) matrix-vector multiplications.
Note that one must re-scale the original matrix as $\tilde H = a(H-b \mathbb 1)$ such that the spectrum of $\tilde H$ is contained in the interval of orthogonality $[-1,1]$ of the Chebyshev polynomials.
Suitable values $a,b \in \mathbb R$ are determined initially with Gershgorin's circle theorem or a few Lanczos sweeps.

From the vectors $|\nu_m\rangle$ 
two scalar products $\eta_{2m} = \langle \nu_m | \nu_m \rangle$, $\eta_{2m+1} = \langle \nu_{m+1} | \nu_m \rangle$ are computed in each iteration step.
Spectral quantities 
are reconstructed from these scalar products in a second computationally inexpensive step,
which is independent of the KPM iteration and needs not be discussed in the context of performance engineering.
For the computation of a spectrally averaged quantity, e.g., the DOS from Eq.~\eqref{DOS}, the trace can be approximated by a sum over several independent random initial vectors as in $\mathrm{tr} [A] \approx (1/R) \sum_{r=1}^R \langle \nu^{(r)}_0 |A|  \nu^{(r)}_0 \rangle$
(see~\cite{Weisse06} for further details).

\begin{figure}[t]  
\begin{algorithmic}  
\For{ $r=0$\ to\ $R-1$ }
\State $| v \rangle  \gets   | \mbox{rand()} \rangle$
\State Initialization steps and computation of $ \eta_0 , \eta_1$  
\For{ $m=1$\ to\ $M / 2$ }
\State $\mbox{swap} ( | w \rangle, | v \rangle )$
\State $\mathrlap{| u \rangle}\phantom{\eta_{2m+1}}  \gets H | v \rangle  $
\Comment{\texttt{spmv()}}
\State $\mathrlap{| u \rangle}\phantom{\eta_{2m+1}}  \gets  | u \rangle - b | v
\rangle   $ \Comment{\texttt{axpy()}}
\State $\mathrlap{| w \rangle}\phantom{\eta_{2m+1}}  \gets -| w \rangle    $
\Comment{\texttt{scal()}}
\State $\mathrlap{| w \rangle}\phantom{\eta_{2m+1}}  \gets  | w \rangle + 2 a |
u \rangle   $ \Comment{\texttt{axpy()}}
\State $\mathrlap{\eta_{2m}}\phantom{\eta_{2m+1}}  \gets   \langle v | v \rangle
$ \Comment{\texttt{nrm2()}}
\State $\eta_{2m+1}  \gets  \langle w | v \rangle  $ \Comment{\texttt{dot()}}
\EndFor
\EndFor
\end{algorithmic}
\caption{Naive version of the KPM-DOS algorithm with corresponding BLAS level
1 function calls. 
Note that the ``swap'' operation is not performed explicitly but merely indicates the logical change of the role of the $v$, $w$ vectors in the odd/even iteration steps.}
\label{alg:kpm_naive}
\end{figure}

A direct implementation of the above scheme
results in the ``naive'' version of the KPM-DOS algorithm shown in Fig.~\ref{alg:kpm_naive}.
One feature of KPM is the very simple implementation of the basic algorithm,
which leaves substantial headroom for performance optimization. 
The above algorithm involves one SpMV and a few BLAS level 1 operations per step.
If only two vectors are stored in the implementation the scalar
products have to be computed before the next iteration step.
That and the multiple individual BLAS level 1 operations on the vectors $|v\rangle, |w\rangle$ call for optimization of the local data access patterns.
A careful implementation reduces the amount of global reductions in the dot products to a
single one at the end of the inner loop.
Furthermore, in its present form the stochastic trace is performed via an outer loop over $R$ random vectors. Although the inner KPM iterations for different initial vectors are independent of each other, 
performance gains compared to the embarrassingly parallel version with $R$
independent runs can be achieved by incorporating the ``trace'' functionality
into the parallelized algorithm. See \cref{sec:performance_parallel} for
detailed performance results.

\section{Algorithm Analysis and Optimization}
\label{sec:algo_optimization}
To get a first overview of the algorithm's properties it is necessary to study
its requirements in terms of data transfers and computational work, both of
which depend on the data types involved. $S_d$ and $S_i$ denote the size of a
single matrix/vector data element and matrix index element, respectively.
$F_a$ ($F_m$) indicates the number of floating point operations (flops) per addition
(multiplication).
\begin{table}[t]
\centering
\begin{tabular}{lrrr}
\hline
Funct. & \# Calls & Min. Bytes/Call & Flops/Call  \\
\hline
\hline
\texttt{spmv()} & $RM/2$ & $N_\mathrm{nz}(S_d+S_i) +$
& $N_\mathrm{nz}(F_a+F_m)$ \\
& & $2NS_d$ & \\
\texttt{axpy()} & $RM$ & $3NS_d$ & $N(F_a+F_m)$ \\
\texttt{scal()} & $RM/2$ & $2NS_d$ & $NF_m$\\ 
\texttt{nrm2()} & $RM/2$ & $NS_d$ &  $N(\lceil F_a/2 \rceil + \lceil F_m/2 \rceil)$ \\
\texttt{dot()} & $RM/2$ & $2NS_d$ &  $N(F_a+F_m)$ \\
\hline
\multirow{2}{*}{KPM} & \multirow{2}{*}{1} & $RM/2[N_\mathrm{nz}(S_d+$ &
$RM/2[N_\mathrm{nz}(F_a+F_m) + $ \\
& & $ S_i) + 13NS_d] $ & $N(\lceil 7F_a/2 \rceil + \lceil 9F_m/2 \rceil)]$ \\
\hline
\end{tabular}
\caption{Minimum number of transferred bytes and executed flops for each
    function involved in \cref{alg:kpm_naive}. }
\label{tab:kpm_analysis}
\end{table}
\Cref{tab:kpm_analysis} shows the minimum number of flops
to execute and memory bytes to transfer for each of the operations involved
in \cref{alg:kpm_naive} and for the entire algorithm.

Generally speaking, algorithmic optimization involves a reduction of resource requirements, 
i.e., lowering either the data traffic or the
number of flops. While we assume the latter to be fixed for this algorithm, it is
possible to improve on the former.
From \cref{alg:kpm_naive} it becomes clear that the
vectors $u$, $v$, and $w$ are read and written several times. An obvious
optimization is to merge all involved operations into a single
one. This is a simple and widely applied code optimization technique, also known
as loop fusion. In our case, we augment the SpMV kernel with the required operations for
shifting and scaling. Furthermore, the needed dot products are being calculated
on-the-fly in the same kernel. 
Note that optimizations of this kind usually require manual implementation
due to the lack of libraries providing exactly the kernel as needed.
The new kernel will be called \texttt{aug\_spmv()}
and the resulting algorithm is shown in \cref{alg:kpm_improved}.
\begin{figure}[bt]
\begin{algorithmic}  
\For{ $r=0$\ to\ $R-1$ }
\State $| v \rangle  \gets   | \mbox{rand()} \rangle$
\State Initialization steps and computation of $ \eta_0 , \eta_1$  
\For{ $m=1$\ to\ $M / 2$ }
\State $\mbox{swap} ( | w \rangle, | v \rangle )$
\State $| w \rangle = 2a(H - b \mathbb 1) | v \rangle -| w \rangle $ \&  \Statex
$\mathrlap{\eta_{2m}}\phantom{\eta_{2m+1}} =
\langle v | v \rangle$ \& \Statex $\eta_{2m+1} = \langle w | v \rangle $
\Comment{\texttt{aug\_spmv()}}
\EndFor
\EndFor
\end{algorithmic}
\caption{Optimization stage 1: Improved version of the KPM-DOS algorithm using the augmented SpMV
kernel, which covers all operations chained by '\&'.}
\label{alg:kpm_improved}
\end{figure}

The data traffic due to the vectors has been reduced
in comparison with the naive implementation in \cref{alg:kpm_naive} by saving 10 vector transfers in each inner iteration. 
A further improvement can be accomplished by exploiting that the same matrix
$H-b \mathbb{1}$ gets applied to $R$ different vectors. By interpreting the vectors as a
single block vector of width $R$, one can get rid of the outer loop and apply
the matrix to the whole block at once. Thus, the resulting operation is an
augmented SpMMV, to be referred
as \texttt{aug\_spmmv()}. The resulting algorithm is shown in \cref{alg:kpm_improved_blocked}. 
\begin{figure}[tb]
\begin{algorithmic}
\State $\mathrlap{| V \rangle}\phantom{| W \rangle} \vcentcolon= | v \rangle_{0..R-1}$ \Comment Assemble vector blocks
\State $| W \rangle \vcentcolon= | w \rangle_{0..R-1}$
\State $\mathrlap{| V \rangle}\phantom{| W \rangle} \gets   | \mbox{rand()} \rangle$
\State Initialization steps and computation of $ \mu_0 , \mu_1$  
\For{ $m=1$\ to\ $M / 2$ }
\State $\mbox{swap} ( | W \rangle, | V \rangle )$
\State $| W \rangle = 2a(H - b \mathbb 1) | V \rangle -| W \rangle $ \&  \Statex
$\mathrlap{\eta_{2m}\mathtt{[:]}}\phantom{\eta_{2m+1}\mathtt{[\;]}} =
\langle V | V \rangle$ \& \Statex $\eta_{2m+1}\mathtt{[:]} = \langle W | V \rangle $
\Comment{\texttt{aug\_spmmv()}}
\EndFor
\end{algorithmic}
\caption{Optimization stage 2: Blocked and improved version of the KPM-DOS algorithm using the augmented SpMMV
kernel. Now, each $\eta$ is a vector of $R$ column-wise dot products of two
block vectors.}
\label{alg:kpm_improved_blocked}
\end{figure}

Now the matrix only has to be read $M/2$ times and the data traffic is 
reduced further. We summarize the data transfer savings
for each optimization stage by showing the evolution of the entire solver's minimum data
traffic $V_\mathrm{KPM}$:
\begin{align}
V_{\mathrm{KPM}} &= RM/2[N_\mathrm{nz}(S_d+S_i) + 13S_dN] \nonumber\\
&\Downarrow \text{Using \PVerb{aug_spmv()}} \nonumber \\
&= RM/2[N_\mathrm{nz}(S_d+S_i) + 3S_dN] \nonumber \\
&\Downarrow \text{Using \PVerb{aug_spmmv()}} \nonumber \\
&= M/2[N_\mathrm{nz}(S_d+S_i) + 3RS_dN]. \label{eq:traffic_kpm_improved_blocked}
\end{align}

It will become evident that data transfers are the bottleneck in this
application scenario; using the relevant data paths to their full
potential is thus the key to best performance. While tasking
approaches for shortening the critical path may seem promising for the
original formulation of the algorithm (see Fig.~\ref{alg:kpm_naive}),
the optimized version in Fig.~\ref{alg:kpm_improved_blocked} is purely
data parallel.

\subsection{General Performance Considerations}
\label{sec:performance}
Using $V_\mathrm{KPM}$ from \cref{eq:traffic_kpm_improved_blocked} and the number of flops as presented in
\cref{tab:kpm_analysis} the minimum code balance of the solver is:
\begin{align}
B_\mathrm{min} &= \frac{N_\mathrm{nz}(S_d+S_i) + 3RS_dN}{R[N_\mathrm{nz}(F_a+F_m) + N(\lceil 7F_a/2 \rceil + \lceil 9F_m/2 \rceil)]} \nonumber \\
&= \frac{N_\mathrm{nzr}/R(S_d+S_i) + 3S_d}{N_\mathrm{nzr}(F_a+F_m) + (\lceil 7F_a/2 \rceil + \lceil 9F_m/2 \rceil)}
\;\frac{\BYTES}{\FLOP}. \nonumber
\end{align}
$N_\mathrm{nzr}=N_\mathrm{nz}/N$ denotes the average number of entries per
row, which is approximately 13 in our test case.
As we are using complex double precision floating
point numbers for storing the vector and matrix data, one data element
requires 16 bytes of storage ($S_d=16$), while 4-byte
integers are used for indexing within the kernels ($S_i=4$).
Note that the code as a whole uses mixed integer
sizes, as 8-byte indices are required for global quantities in large-scale runs.
Furthermore, for complex arithmetic it holds that $F_a=2$ and $F_m=6$. 
Using the actual values for the test problem, we arrive at
\begin{align}   
B_\mathrm{min}(R) &= \frac{13/R(16+4) + 3\cdot16}{13(2+6) + (\lceil 7\cdot2/2
\rceil + \lceil 9\cdot6/2 \rceil)}
\;\frac{\BYTES}{\FLOP} \nonumber \\
&= \frac{260/R+48}{138}\; \frac{\BYTES}{\FLOP} \label{eq:codebalance_topi}\\
B_\mathrm{min}(1) &\approx 2.23\; \frac{\BYTES}{\FLOP}
\label{eq:codebalance_topi_R1} \\
\lim_{R \to \infty} B_\mathrm{min} &\approx 0.35\; \frac{\BYTES}{\FLOP}
\label{eq:codebalance_topi_Rinf}
\end{align}
Usually the actual code balance is larger than $B_\mathrm{min}$. This is mostly
due to part of the SpM(M)V input vector being read from main
memory more than once. This can be caused by an unfavorable matrix sparsity
pattern or an undersized last level cache (LLC). We quantify the performance impact 
by a factor $\Omega=V_\mathrm{meas}/{V_\mathrm{KPM}}$, with
$V_\mathrm{meas}$ being the actual data transfer volume in bytes as measured with, e.g.,
LIKWID~\cite{Likwid} on CPUs and with NVIDIA's \texttt{nvprof} ~\cite{nvprof} profiling tool on
NVIDIA GPUs. Thus, the actual code balance is
\begin{equation}
B=\Omega B_\mathrm{min}\;.
\end{equation}

Following the ideas of Gropp et al.~\cite{Gropp99} and Williams et al.~\cite{Williams09}, 
a simple roof\/line model can be constructed. The roof\/line model assumes that an
upper bound for the achievable performance of a loop with code balance $B$ can be
predicted as the minimum of the theoretical peak performance $P^\mathrm{peak}$ and the
performance limit due to the memory bandwidth $b$:
\begin{equation}
    P^* = \mathrm{min}\left(P^\mathrm{peak},\frac{b}{B}\right).
\label{eq:roofline}
\end{equation}
The large code balance for $R=1$ (\cref{eq:codebalance_topi_R1}) indicates that the kernel will be
memory-bound in this case on modern standard hardware, i.e., the maximum
memory-bound performance according to \cref{eq:roofline} is
\begin{equation}
    P^*_\mathrm{MEM} = \frac{b}{B}.
\label{eq:roofline_mem}
\end{equation}

An important observation from \cref{eq:codebalance_topi_R1,eq:codebalance_topi_Rinf} is
that the code balance decreases when $R$ increases, i.e., when
substituting SpMV by SpMMV. In other words,
the kernel execution becomes more and more independent from the
original bottleneck. On the other hand, 
larger vector blocks require more space in the
cache which may cause an increase of $\Omega$ and, consequently, the code balance.
See ~\cite{Roehrig14} for a more detailed analysis of this effect.
The application of the roof\/line model will be discussed in
\cref{sec:perfmodel_CPU}.

\section{Testbed and Implementation}
\label{sec:testbed}
\Cref{tab:architectures} shows relevant architectural properties
of the benchmark systems.
\begin{table}[t]
\centering
\begin{tabular}{lcccccc}
\hline
 & Clock & SIMD & Cores/ & $b$ & LLC & $P^{\mathrm{peak}}$ \\
 & (MHz) & (Bytes) & SMX & (GB/s) & (MiB) & (Gflop/s) \\
\hline
\hline
IVB & 2200 & 32 & 10 & 50 & 25 & 176 \\
SNB & 2600 & 32 & 8 & 48 & 20 & 166.4 \\
K20m & 706 & 512 & 13 & 150 & 1.25 &  1174 \\ 
K20X & 732 & 512 & 14 & 170 & 1.5 & 1311 \\
\hline
\end{tabular}
\caption{Relevant properties of all architectures used in this paper: 
Intel Xeon E5-2660 v2 (``IVB'') with fixed clock frequency, Intel Xeon E5-2670
(``SNB'') with turbo mode enabled, NVIDIA Tesla K20m
with ECC disabled, and NVIDIA Tesla K20X with ECC enabled} 
\label{tab:architectures}
\end{table}
Simultaneous multithreading (SMT) has been enabled on the CPUs, which results in a total thread count of twice
the number of cores. Both GPUs implement the ``Kepler'' architecture where each
Streaming Multiprocessor (SMX) features 64 double precision units capable of
fused multiply add (FMA).
The Intel C Compiler (ICC) version 14 has been used for the CPU
code. For the GPU code, the CUDA toolkit 5.5 was employed.

Measurements for the node-level performance analysis
(\cref{sec:perfmodel_CPU,sec:perfmodel_GPU}) have been conducted on the
\emph{Emmy}\footnote{https://www.hpc.rrze.fau.de/systeme/emmy-cluster.shtml}
cluster at Erlangen Regional Computing Center (RRZE). This cluster contains a
number of nodes combining two IVB CPUs with two K20m GPUs.

For large-scale production runs we used the heterogeneous petascale cluster
\emph{Piz Daint}\footnote{http://www.cscs.ch/computers/piz\_daint/index.html},
a Cray XC30 system located at the Swiss National Computing Centre (CSCS) in Lugano, Switzerland.
Each of this system's 5272 nodes consists of one SNB CPU and one K20X GPU.


\subsection{General Notes on the Implementation}
\label{sec:implementation}
Although the compute platforms used in this work are heterogeneous at first sight, they
have architectural similarities which enable optimization
techniques that are beneficial on both architectures. 
An important property in this regard is data
parallelism.

Modern CPUs feature \emph{Single Instruction Multiple Data} (SIMD) units which
enable data-parallel processing on the core level. The current Intel CPUs used here
implement the AVX instruction set, which contains 256-bit wide
vector registers.  Hence, four real or two complex numbers 
can be processed at once in double precision. 

The equivalent hardware feature on GPUs is called \emph{Single Instruction
Multiple Threads} (SIMT), which can be seen as ``SIMD on a thread
level~\cite{Volkov08}.'' Here, a
group of threads called \emph{warp} executes the same instruction at a time. On
all modern NVIDIA GPUs a warp consists of 32 threads, regardless of the data
type. Instruction divergence within a warp causes serialization of the thread
execution.
Up to 32 warps are grouped in a \emph{thread block}, which is
the unit of work scheduled on an SMX.

For an efficient utilization of SIMD/SIMT processing the data access has to
be contiguous per instruction. On GPUs, \emph{load coalescing} (i.e.,
subsequent threads have to access subsequent memory locations) is crucial
for efficient global load instructions. Achieving efficient SIMD/SIMT
execution for SpMV is connected to several issues like zero fill-in and the need
for gathering the input vector data~\cite{Kreutzer14}. For SpMV, vectorized
access can only be achieved with respect to the matrix data. However, in the case of
SpMMV this issue can be solved since contiguous data access is possible across the vectors.
Note that it is necessary to store the vectors in an interleaved way (row-major)
for best efficiency. If this is not compatible with the data layout of the
application, transposing the block vector data may be required.
Vectorizing the right-hand side vector data access has the convenient
advantage that matrix elements can be
accessed in a serial manner, which eliminates the need for any special
matrix format. Hence, the CRS format (similar to SELL-1) can be used
on both architectures without drawbacks.  
It is worth noting that CRS/SELL-1 may yield even better SpMMV performance
than a SIMD-aware storage format for SpMV like SELL-32, because 
matrix elements within a row are stored consecutively.

\subsection{CPU Implementation}
The CPU kernels have been hand-vectorized using AVX compiler intrinsics. A
custom code generator was used to create fully unrolled versions of the kernel
codes for different combinations of the SELL chunk height and the block vector width.  As
the AVX instruction set supports 32-byte SIMD units, a minimal vector block
width of two is already sufficient for achieving perfectly vectorized access to
the (complex) vector data.

For memory-bound algorithms (like the naive implementation in
\cref{alg:kpm_naive}) and large working sets, efficient vectorization may
not be required for optimal performance. However, as discussed in
\cref{sec:performance}, our optimized kernel is
no longer strictly bound to memory bandwidth. Hence, efficient vectorization is a crucial
ingredient for high performance. To guarantee best results,
manual vectorization cannot be avoided, especially in case of complex arithmetic. 

\subsection{GPU Implementation}
\label{sec:implementation_GPU}
\begin{figure*}[tb]
\centering
\includegraphics[width=\textwidth]{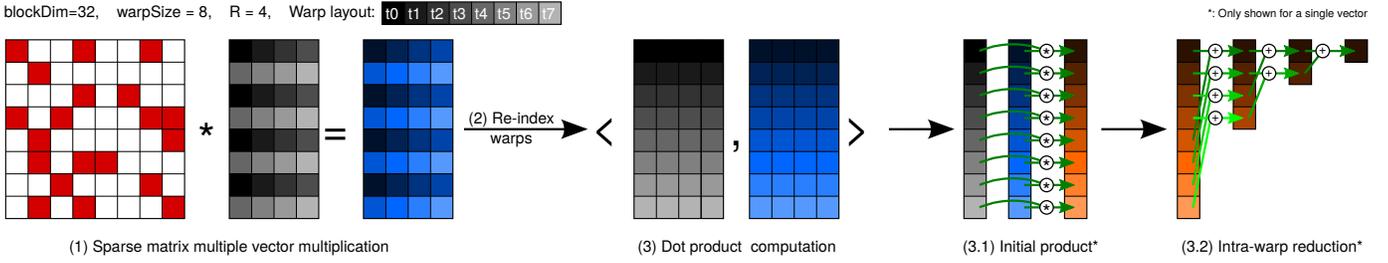}
\caption{GPU implementation of SpMMV with on-the-fly dot product. Only a single
thread block is shown.}
\label{fig:cuda_spmv_dot}
\end{figure*}
The GPU kernels have been implemented using CUDA and hand-tuned for the Kepler
architecture. There are well-known approaches for efficient SpMV 
(see, e.g., ~\cite{Bell09}, \cite{Kreutzer14} and references therein),
but the augmented SpMMV kernel requires more effort. 
In particular for the implementation of on-the-fly dot product
computations a sensible thread management is crucial. 
\Cref{fig:cuda_spmv_dot} shows how the threads are mapped to the
computation in the full \texttt{aug\_spmmv()} kernel. For the sake of easier
illustration, relevant architectural properties have been set to smaller values.
In reality, the \texttt{warpSize} is 32 on Kepler and the maximum (and also the
one which is used)
\texttt{blockDim} is 1024. The threads in a warp are colored with increasing
brightness in the figure.
Note that this implementation is optimized towards relatively large vector
blocks ($R \gtrsim 8$). In the following we explain the components of the augmented
SpMMV kernel.

\subsubsection{SpMMV}

The first step of the kernel is the SpMMV. In order to have coalesced access to
the vector data, the warps must be arranged along block vector rows. Obviously,
perfectly coalesced access can only be achieved for block vector widths which are at least
as large as the warp size. We have observed, however, that smaller block
widths achieve reasonable performance as well. In \cref{fig:cuda_spmv_dot}
the load of the vector data would be divided into two loads using half a warp
each.

\subsubsection{Re-index warps} 

Operations involving reductions are usually problematic in GPU programming
for two reasons: First, reductions across multiple blocks require thread
synchronization among the blocks. Second, the reduction inside a block
demanded the use of shared memory on previous NVIDIA architectures;
this is no longer true for Kepler, however. This architecture implements \emph{shuffle} instructions,
which enable sharing values between threads in a warp without having to use
shared memory~\cite{KeplerWhitepaper}. For the dot product computation, the values which have to be
shared between threads are located in the same vector (column of the block).
This access pattern is different from the one used in step (1), where subsequent threads access different
columns of the block. Hence, the thread indexing in the warps has to
be adapted. Note that no data actually gets transposed but merely the indexing
changes.

\subsubsection{Dot product}

The actual dot product computation consists of two steps. 
Computing the initial product is trivial, as each thread only computes the
product of the two input vectors.
For the reduction phase, subsequent invocations of the shuffle instruction as implemented in the Kepler
architecture are used. In total, $\mbox{log}_2(\texttt{warpSize})$ reductions are required
for computing the full reduction result, which can then be obtained from the first thread. For the final
reduction across vector blocks, CUB~\cite{cub} has been used
(not shown in \cref{fig:cuda_spmv_dot}).

\section{Performance Models}
In this section we apply the analysis from \cref{sec:performance} to both our
CPU and GPU implementation using an IVB CPU and a K20m GPU.
We use a domain of size $100\times100\times40$ if not stated otherwise.
This results in a matrix with $1.6\cdot 10^6$ rows. Thus, neither the matrix nor the
vectors fit into any cache on either architecture.

\subsection{CPU Performance Model}
\label{sec:perfmodel_CPU}
\begin{SCfigure}[][tb]
\centering
\caption{Socket scaling on IVB. The roof\/line prediction results from the
measured attainable bandwidth $b$
from \cref{tab:architectures}, the code balance $B_\mathrm{min}$ from \cref{eq:codebalance_topi_R1}, and
$\Omega=1$ (best case) in \cref{eq:roofline}}
\includegraphics[width=.5\columnwidth]{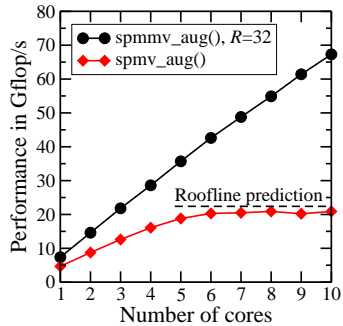}
\label{fig:socket_scaling}
\end{SCfigure}

The relevant architectural bottleneck for SpM(M)V changes when increasing the
block vector width.
This assertion is confirmed by the intra-socket scaling
performance (see \cref{fig:socket_scaling}) on IVB.
The performance of the SpMV kernel is clearly bound by main memory bandwidth,
saturating at a level (dashed line) which is
reasonably close the roof\/line prediction obtained from \cref{eq:roofline_mem}.
In contrast, the SpMMV kernel performance scales almost linearly within a socket. 
This indicates that
the relevant bottleneck is either the bandwidth of some cache level or the
in-core execution. It turns out that taking into account the L3 cache yields sufficient insight 
for a qualitative analysis of the performance bottlenecks (for recent work on refined 
roof\/line models for SpMMV see ~\cite{Aktulga14} where both the L2
and L3 cache were considered). 
The roof\/line model (\cref{eq:roofline}) can be modified by defining a more precise
upper performance bound than $P^\mathrm{peak}$ for loops that are decoupled from 
memory bandwidth: 
\begin{equation}
    P^* = \mathrm{min}(P^*_\mathrm{MEM},P^*_\mathrm{LLC}).
\label{eq:roofline_custom}
\end{equation}
Here, $P^*_\mathrm{LLC}$ is a performance limit for the last level cache,
which is determined through benchmarking a
down-sized problem where the whole working set (matrix and vectors) fits into
the L3 cache of IVB 
while keeping the matrix as similar as
possible to the memory-bound test case.
\begin{figure}[b]
\centering
\includegraphics[width=\columnwidth]{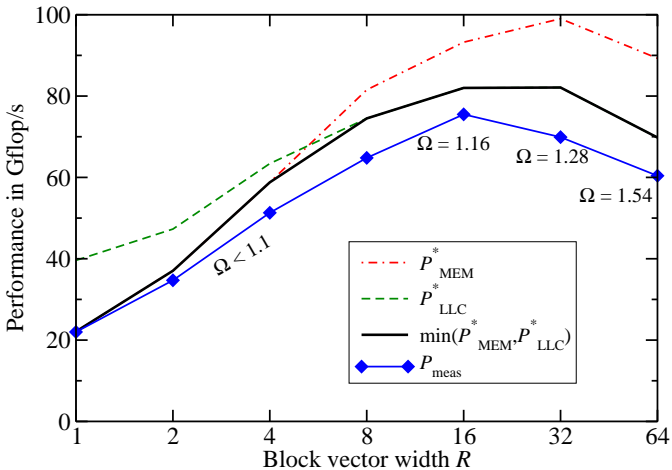}
\caption{Custom roof\/line model for the augmented SpM(M)V kernel on IVB. Upper
bounds for memory- and L3-boundedness, as well as the minimum of both, can be
seen together with measured performance data.}
\label{fig:cpu_perfmodel}
\end{figure}
A comparison of our custom roof\/line model with measured performance 
for the augmented SpM(M)V kernel is shown in \cref{fig:cpu_perfmodel}.
The shift of the relevant bottleneck can be identified:
For small $R$ the kernel is memory-bound and the performance can be
predicted by the standard roof\/line model (\cref{eq:roofline}) 
with high accuracy. At larger $R$,
the kernel's execution decouples from main memory. A high
quality performance prediction is more complicated in this region,
but our refined model (\cref{eq:roofline_custom}) does not deviate
by more than 15\% from the measurement.
A further observation from \cref{fig:cpu_perfmodel} is the impact of $\Omega$
(see annotations in the figure)
on the code balance and on $P^*_\mathrm{MEM}$: For large $R$ the maximum
achievable performance decreases although the minimum code balance (see
\cref{eq:codebalance_topi}) originally suggests otherwise.

\subsection{GPU Performance Model}
\label{sec:perfmodel_GPU}
\begin{SCfigure}[][bt]
\centering
\caption{Measured data volume per block vector 
for different memory system components on the Kepler 
GPU running the simple SpMMV kernel.}
\includegraphics[width=.5\columnwidth]{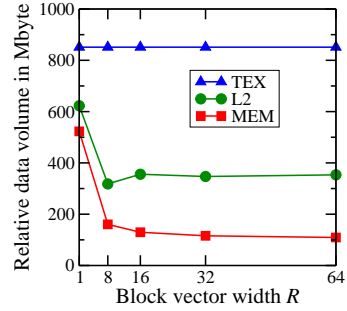}
\label{fig:gpu_perfmodel_vol}
\end{SCfigure}

On the GPU, establishing a custom roofline model as in \cref{eq:roofline_custom}
is substantially more difficult because one can not use the GPU to full
efficiency with a data set that fits in the L2 cache.
Hence, the performance model for the GPU will be more of a qualitative nature.
The Kepler architecture is equipped with two caches that are relevant for the
execution of our kernel. Information on these caches can be found in 
~\cite{KeplerWhitepaper} and ~\cite{KeplerTuningGuide}:
\begin{enumerate}
\item \textit{L2 cache:}
The L2 cache is shared between all SMX units. In
the case of SpMV, it serves to alleviate the penalty of unstructured accesses to
the input vector.
\item \textit{Read-only data cache:}
On Kepler GPUs there is a 48\,KiB read-only data cache (also called texture cache)
on each SMX. This cache has relaxed memory coalescing
rules, which enables efficient broadcasting of data to
all threads of a warp. It can be used in a transparent way if read-only data
(such as the matrix and input vector in the \texttt{aug\_spmmv()} kernel)
is marked with both the \texttt{const} and \texttt{\_\_restrict\_\_} qualifiers.
In the SpMMV kernel, each  matrix entry needs to be broadcast to the threads of a
warp (see \cref{sec:implementation_GPU} for details), which makes this kernel a
very good usage scenario for the read-only data cache.
\end{enumerate}
In \cref{sec:implementation_GPU} we have described how the computation of dot
products complicates the augmented SpMMV kernel. For our bottleneck
analysis we thus consider the plain SpMMV kernel, the augmented SpMMV kernel (but
\emph{without} on-the-fly computation of dot products), and finally the
full augmented SpMMV kernel.
To quantify the impact of different memory system components we present the measured
data volume when executing the simple SpMMV kernel (the qualitative
observations are similar for the other kernels) for each of them in \cref{fig:gpu_perfmodel_vol}.
The data traffic coming from the texture cache scales linearly with $R$ because
the scalar matrix data is broadcast to the threads in a warp via this cache. 
The accumulated data volume across all hierarchy levels decreases for increasing
$R$, which is due to the shrinking relative impact of the matrix on the data
traffic. A potential further reason for this effect is higher load efficiency in
the large $R$ range.

\begin{figure}[tb]
\centering
\includegraphics[width=\columnwidth]{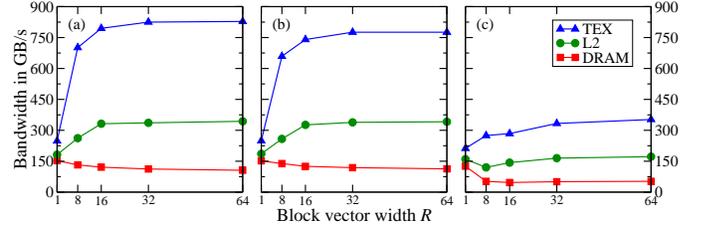}
\caption{Measured bandwidth on the K20m GPU for different memory system components and
kernels: (a) simple SpM(M)V kernel, (b) augmented SpM(M)V kernel without on-the-fly dot product
computation, and (c) fully augmented SpM(M)V kernel.}
\label{fig:gpu_perfmodel_bw}
\end{figure}
\Cref{fig:gpu_perfmodel_bw} shows DRAM, L2 cache, and Texture cache
bandwidth measurements for the three kernels
mentioned above. 
At $R=1$ the 
DRAM bandwidth is around 150 GB/s for the first two kernels, which is equal to the maximum attainable
bandwidth on this device (see \cref{tab:architectures}; as expected, the kernel is memory bound. 
The bandwidths drawn from L2 and Texture cache are not much higher
than the DRAM bandwidth in this case.
With growing $R$ the DRAM bandwidth decreases 
while the bandwidths of L2 and Texture cache increase and eventually saturate.
Thus, the relevant bottleneck is changed from DRAM to cache bandwidth as
the computational intensity of the kernel goes up. For the fully augmented
SpM(M)V kernel (right panel in \cref{fig:gpu_perfmodel_bw}), the qualitative
curve shapes are similar to the other two kernels but all measured bandwidths
are at a significantly lower level. This is caused by the dot product computation
with all its issues (cf.~\cref{sec:implementation_GPU}), making 
instruction latency the relevant bottleneck. However, this kernel still yields significantly
higher performance than an implementation with separate dot product computation.

All these observations and conclusions coincide with
the bottleneck analysis of the NVIDIA Visual Profiler. For all kernels it
determines the DRAM bandwidth as the relevant bottleneck at $R=1$. 
At larger $R$
the L2 cache bandwidth is the bottleneck for the kernels without on-the-fly dot
product calculations. Otherwise, i.e., when including dot products, the reported
bottleneck is latency.

\section{Performance Results}
\label{sec:performance_results}

\subsection{Symmetric Heterogeneous Execution}
\label{sec:implementation_heterogeneous}
MPI is used as a communication layer for heterogeneous execution, and the
parallelization across devices is done on a per-process basis. On a single
heterogeneous node, simultaneous hybrid execution could also be implemented
without MPI. However, using MPI already on the node level enables easy scaling
to multiple heterogeneous nodes and portability to other heterogeneous
systems. We use one process for each CPU/GPU in a node and OpenMP
within CPU processes.  A GPU process needs a certain amount of
CPU resources for executing the host code and calling GPU kernels, for which
one CPU core is usually sufficient. Hence, for the heterogeneous measurements in this
paper one core per socket was ``sacrificed'' to its GPU.
Each process runs in its own disjoint CPU set, i.e., there are no resource
conflicts between the processes on a node. 

The assembly of communication buffers
in GPU processes is done in a GPU kernel. Only the elements which need to be
transferred are copied to the host side before sending them to the
communication partners. This is done
via pinned memory in order to achieve a high transfer rate.

An intrinsic property of heterogeneous systems is that the components
usually do not only differ in architecture but also in
performance. For optimal load balancing this difference has to be
taken into account for work distribution. In our execution environment
a weight has to be provided for each process.  From this weight we compute
the amount of matrix/vector rows that get assigned to it.

\subsection{Node-Level Performance}
\label{sec:performance_node}
\begin{SCfigure}[][bt]
\centering
\caption{Node-level performance for each
optimization stage on Piz Daint. 
The parallel efficiency of the heterogeneous execution is shown on top of the heterogeneous
performance bars.}
\includegraphics[width=.5\columnwidth]{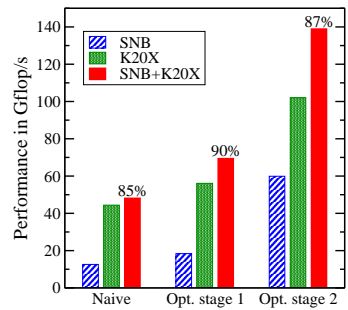}
\label{fig:node_performance}
\end{SCfigure}

\Cref{fig:node_performance} shows the performance on a heterogeneous node for
both architectures and all optimization stages. Single architectures solve for a
$200\times100\times40$ domain, and a $400\times100\times40$ domain has been used for the
heterogeneous runs.
 All weights have been tuned
experimentally. However, a good guess is to calculate the weights from the
single-device performance numbers. The maximum speed-up which can be achieved on a single node,
i.e., the speed-up between the naive CPU-only implementation and the
fully optimized 
heterogeneous version, is more than a factor of 10.  However, a more realistic
usage model of a GPU-equipped node is the naive GPU-only variant.
Here, a speed-up of 2.3$\times$ can be achieved by algorithmic optimizations and
careful implementation.  On top of that, another 36\% can be gained by
enabling fully heterogeneous execution including the CPU.  The parallel
efficiency of the heterogeneous implementation with respect to the sum of the
single-architecture performance levels tops out at 85--90\%. The gap to optimal
efficiency has two major reasons:
First, the heterogeneous implementation includes communication over the relatively
slow PCI Express bus. Second, one CPU core is used for GPU management. As the CPU
kernel's bottleneck is not memory bandwidth, excluding one core from the
computation results in a performance decrease on the CPU side.

\subsection{Large-Scale Parallel Performance}
\label{sec:performance_parallel}
\begin{figure}[tb]
\centering
\includegraphics[width=\columnwidth]{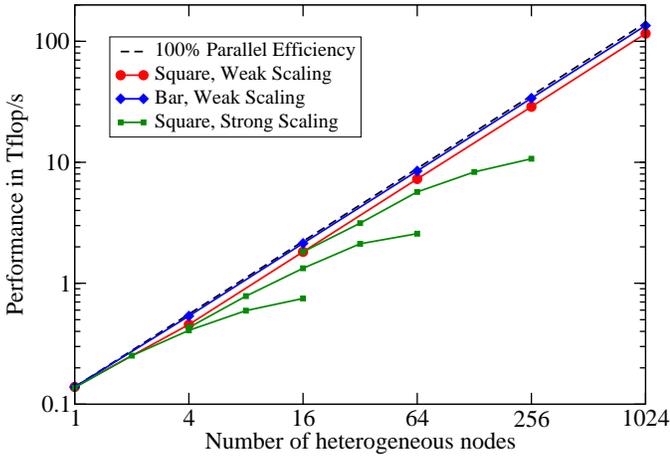}
\caption{Weak scaling performance of the full KPM solver on Piz Daint for the
    ``Square'' and ``Bar'' test cases. The strong
scaling curves show the performance for the ``Square'' case at a problem size 
as defined by the first point of each curve.}
\label{fig:scaling_perf}
\end{figure}
\Cref{fig:scaling_perf} shows scaling data on up to 1024 nodes of Piz
Daint for the topological material application scenario. For weak scaling 
we solve for two different domains: First, we consider
a tile with fixed height $N_z=40$ and equally growing width and length
(``Square''). The second test case represents a domain with fixed width
$N_y=100$ and height $N_z=40$, and growing $N_x$ (``Bar''). For both cases, the
baseline performance on a single node corresponds to the same system as 
in \cref{fig:node_performance}, i.e., a domain of size $400\times100\times40$. In the
``Square'' test, the $y$ dimension increases to 400 when going to four nodes in
order to have a quadratic tile. The drop in parallel efficiency in this region
is a result of the growing number of processors in the $y$ direction, which
leads to an increase in communication volume. 
On larger node counts the number of nodes quadruples in
each step while the extent in $x$ and $y$ direction doubles. In the ``Bar'' test,
the $x$ dimension increases by 400 for each added node. 
The strong scaling curves always represent the performance for
a fixed problem size as given by the data set at the first point of each curve.
The largest system solved in these runs is described by a  matrix with over $6.5\cdot 10^9$ rows. 

Looking at the non-blocked version of the algorithm (\cref{alg:kpm_improved}),
one may argue that there is no dependency between outer loop iterations, and
highly efficient parallelization should be easily achieved by just running $R$ 
instances of the loop code. However, our optimization stage 2 has shown that  
it is just the incorporation of the $R$ loop that enables the algorithm to decouple
from the memory bandwidth; solving the problem in ``throughput mode''
will thus incur a significantly higher overall cost. We illustrate this
difference in \cref{tab:largescale_comparison}, which 
summarizes the resource requirements of three
different variants to solve the largest problem: the augmented SpMV from 
\cref{alg:kpm_improved}, the augmented SpMMV with a global reduction 
over dot products in each iteration (cf. \cref{sec:kpm}), and the final optimized version
with a single global reduction at the end.
\begin{table}[bt]
\centering
\begin{tabular}{lrrr}\hline
Version & Tflop/s & Nodes & Node hours \\
\hline
\hline
\texttt{aug\_spmv()} & 14.9 & 288 & 164 \\
\texttt{aug\_spmmv()}$^*$& 107 & 1024 & 81 \\
\texttt{aug\_spmmv()} & 116 & 1024 & 75 \\
\hline
\end{tabular}
\caption{Overview of required resources for solving the largest system with
$R=32$ and $M=2000$. The non-blocked version \texttt{aug\_spmv()} has been run
in throughput mode. \texttt{aug\_spmmv()}$^*$ indicates a version where a
global reduction over the dot products has been done in each iteration instead
of once at the very end.\label{tab:largescale_comparison}}
\end{table}
The data shows impressively that the embarrassingly $R$-parallel version
is more than a factor of two more expensive in terms of compute
resources (node hours) than the optimal version. Reducing the number of global
reductions increases the performance by 8\%. Note that this factor strongly
depends on the communication patterns and can be substantially higher for other
matrices.


\section{Conclusion and Outlook}
In this work we have performed systematic, model-guided performance engineering
for the KPM-DOS algorithm leading to a substantial increase in node-level
performance on the CPU and the GPU. This was achieved by crafting a
problem-specific loop kernel that has all required operations fused in and
achieves minimal theoretical code balance. The performance analysis of the
optimized algorithm on the CPU and on the GPU revealed that the optimizations
led to a complete decoupling from the main memory bandwidth for relevant
application cases on both the CPU and the GPU. Finally we have embedded our
optimized node-level kernel into a massively parallel, heterogeneous application
code. For the interesting application scenario of topological insulators we have
demonstrated the scalability, performance, and resource efficiency of the
implementation on up to 1024 nodes of a petascale-class Cray XC30 system.  All
software which has been developed within the scope of this work is available for
download~\cite{GHOST}.

In the future we will apply our findings and code to other blocked sparse linear
algebra algorithms besides KPM.
Several open questions remain regarding possible improvements of
our approach.  A future step could be to determine the process weights for
heterogeneous execution automatically and take this burden away from the user.
Furthermore, heterogeneous MPI communication is a field which has room for
improvement.  A promising optimization is to establish a pipeline for this
GPU-CPU-MPI communication, i.e., download parts of the communication buffer to
the host and transfer previous chunks via the network at the same time.  It will
also be worthwhile investigating further optimization techniques such as cache
blocking~\cite{Im04} for the CPU implementation of SpMMV.
Although the Intel Xeon Phi coprocessor is already supported in our software, we
still have to carry out detailed model-driven performance engineering for this
architecture and the KPM application.

\section*{Acknowledgments}
We are indebted to the Swiss National Computing Centre for granting access to
Piz Daint. This work was supported (in part) by the
German Research Foundation (DFG) through the Priority Programs 1648
``Software for Exascale Computing'' under project ESSEX
and 1459 ``Graphene''.

\bibliographystyle{IEEEtran}
\bibliography{ipdps15}

\end{document}